\documentstyle[12pt,epsf,epsfig,wrapfig,rotate,graphics]{article}
\begin{document}

\begin{center}
{\bfseries SPIN EFFECTS IN ELASTIC BACKWARD P-D SCATTERING}

\vskip 5mm
A.P.Ierusalimov$^{1 }$, \underline{G.I.Lykasov }$^{1 \dag}$ and
M.Viviani$^{2}$

\vskip 5mm
{\small
(1) {\it
Joint Institute for Nuclear Research, Dubna, 141980, Moscow region,
Russia}\\
(2) {\it
INFN, Sezione di Pisa, Largo Bruno Pontecorvo, I-56127, Pisa, Italy}\\
$\dag$ {\it
E-mail:lykasov@jinr.ru}}
\end{center}

\vskip 5mm
\begin{abstract}
The elastic backward proton-deuteron scattering is analyzed including 
both relativistic effects in the deuteron and the reaction mechanism.
It is shown that inclusion of the graphs corresponding to the emission, 
rescattering and absorption of the virtual pion by a deuteron nucleon 
in addition to the one-nucleon exchange graph allows a rather satisfactory 
description of all the experimental data on the differential cross section, 
tensor analyzing power of the deuteron and transfer polarization in this 
reaction .
\end{abstract}

\vskip 8mm 

\section{Introduction}
As is known, the study of polarization phenomena in hadron and hadron-nucleus 
interactions gives more detailed information on dynamics of their interactions
and the structure of colliding particles. 
The elastic backward proton-deuteron scattering has been experimentally and 
theoretically studied in Saclay \cite{Saclay}, Dubna and at the JLab (USA) 
\cite{Azhgirei,Punjabi}. 
Up to now all these data cannot be described within the one-nucleon exchange 
model (ONE) including even the relativistic effects in the deuteron \cite{KermKissl:1969,
BagGross:1969,IL:2001}.  

In this paper we analyze the elastic backward proton-deuteron scattering within the 
relativistic approach including the ONE and the high order graphs corresponding to 
the emission, rescattering and absorption of the virtual pion by a deuteron nucleon.

\section{One-nucleon exchange model}
The studies of the elastic backward proton-deuteron scattering within the nonrelativistic
ONE and the relativistic invariant one-nucleon exchange model (RONE)
are presented in Ref.\cite{KermKissl:1969} and Ref.\cite{BagGross:1969} respectively.
The differential cross section calculated within the RONE (Fig.1a) can be presented in the following 
form \cite{IL:2001}:     
\begin{equation}
\frac{d\sigma}{d\Omega}|_{c.m.s.}=\frac{6\pi^2}{s}m^2(m^2-u)^2\mid\Psi_d(q_s^2)\mid^4~,
\end{equation}
where $\Psi_d(q_s^2)$ is the deuteron wave function; $s$ is the square of the initial energy
in the $p-D$ c.m.s., $u$ is the square of momentum transfer 
from initial deuteron to final proton; $q_s^2=\frac{1}{4}s_{12}-m^2$, $s_{12}=(k_1+k_2)^2$, 
$k_1,k_2$ are the four-momenta of neutron and proton in the deuteron, $m$ is the nucleon mass.
Unfortunately, the ONE and the RONE do not allow a satisfactory description of all the observables
at the kinetic energy of backward scattered protons $T_p>0.6$ GeV \cite{IL:2001} . 

\section{One-nucleon and one-pion exchange graphs}
As was shown in Ref.\cite{Barry:1972}, the contribution of the high-order 
graphs in the $p-D$ backward elastic scattering corresponding to the
emission, rescattering and absorption of the virtual pion by a deuteron
nucleon can be sizable at initial energies corresponding
to possible creation of the $\Delta$-isobar at the $\pi-N$ vertex, see Fig.1c. 
The corrections to the ONE graph of Fig.1a were also analyzed in other papers,
see for example Ref.\cite{Uzikov} and references therein. As was shown in 
Refs.\cite{DL:1990,DN:1989} the contribution of the one-pion exhange graphs to the 
deuetron stripping reaction of type $D+p\rightarrow p+X$ can be also sizable at
the initial energies close to a possible $\Delta$-isobar creation in the 
intermediate state.
 \begin{figure}[htb]
\epsfig{file=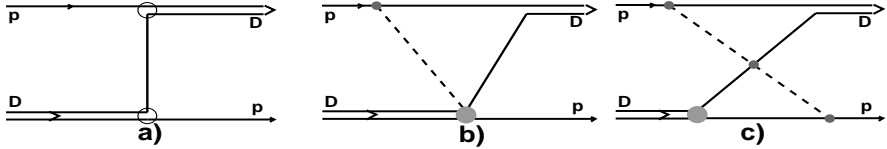,height=2.cm,width=12.cm }
\caption[Fig.1]{One-nucleon exchange graph (a) one-pion exchange graph for the process
$p+D\rightarrow D+p$ (b), and its equivalent graph (c).} 
\end{figure}

\section{Results and discussion}
We calculated the center-of-mass differential cross section, the tensor analyzing
power of the deuteron $T_{20}$ and the transfer polarization $\kappa_0$ in the
elastic backward $p-D$ squattering including the RONE graph (Fig.1a) and the graphs 
of Fig.1c. These results are presented in Figs.(2,3). In Figs.(2,3) curves 
$1$ and $3$ correspond to the total calculation and the RONE computation using 
the Reid soft core deuteron wave function, whereas the lines $2$ and $4$ are the same 
calculations but for the Argon-18 $N-N$ potential. As is evident from Figs.(2,3) the 
RONE  allows us to describe $d\sigma/d\Omega_{c.m.s.}$ and $T_{20}$ at initial deuteron momenta 
up to~$1.5 GeV/c$, whereas the transfer polarization $\kappa_0$ is not described within the RONE 
in the wide interval of deuteron momenta $1.(GeV/c)<p_d^{l.s.}<4.(GeV/c)$. Figures.(2,3) show 
that the total calculation of all the observables including the graphs of Fig.1a and Fig.1c 
results in a rather satisfactory description of the experimental data . 
The graphs of Fig.1c were calculated using the monopole form factor for the virtual pion with the
cut-off parameter about $1.GeV/c$. The $\pi-N$ amplitude entering into the  $\pi-N$
vertex of the Fig.1c graph was taken from the $\pi-N$ phase shift analysis.

One can conclude that the calculation of all the observables for the elastic backward $p-D$
scattering within the relativistic invariant approach including the RONE graphs and the one-pion
exchange graphs of Fig.1c type results in a rather satisfactory description of the experimental 
data at initial deuteron momenta up to $7 GeV/c$. Note that we do not include the six-quark
admixture in the deuteron wave function. This effect can probably be important at larger 
initial momenta because the contribution of the Fig.1c graphs decreases when $p_d^{l.s}$
increases, as is shown in Fig.2.
   
\begin{center}
\begin{figure}[htb]
\includegraphics[angle=-90,width=0.6\textwidth]{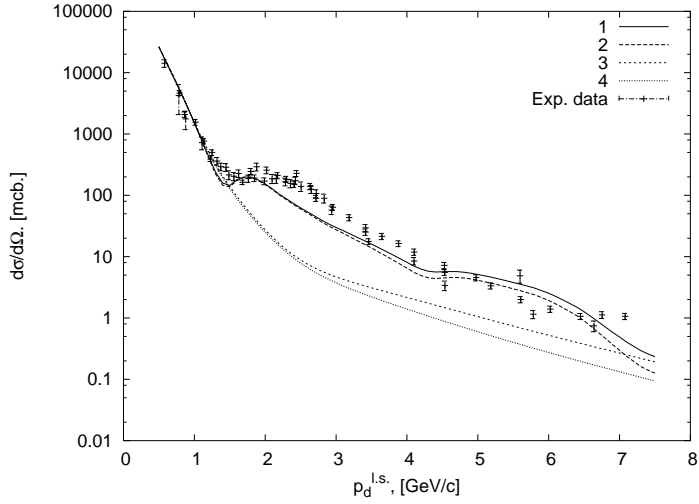}
\caption[Fig.2]{The center-of-mass differential cross section  
$d\sigma/d\Omega_{c.m.s.}$ for the elastic backward $p-D$ scattering  
as a function of the deuteron momentum $p_d^{l.s}$ in the laboratory system.} 
\end{figure}
\end{center}

\begin{figure}[t]
\rotatebox{270}%
{\psfig{file=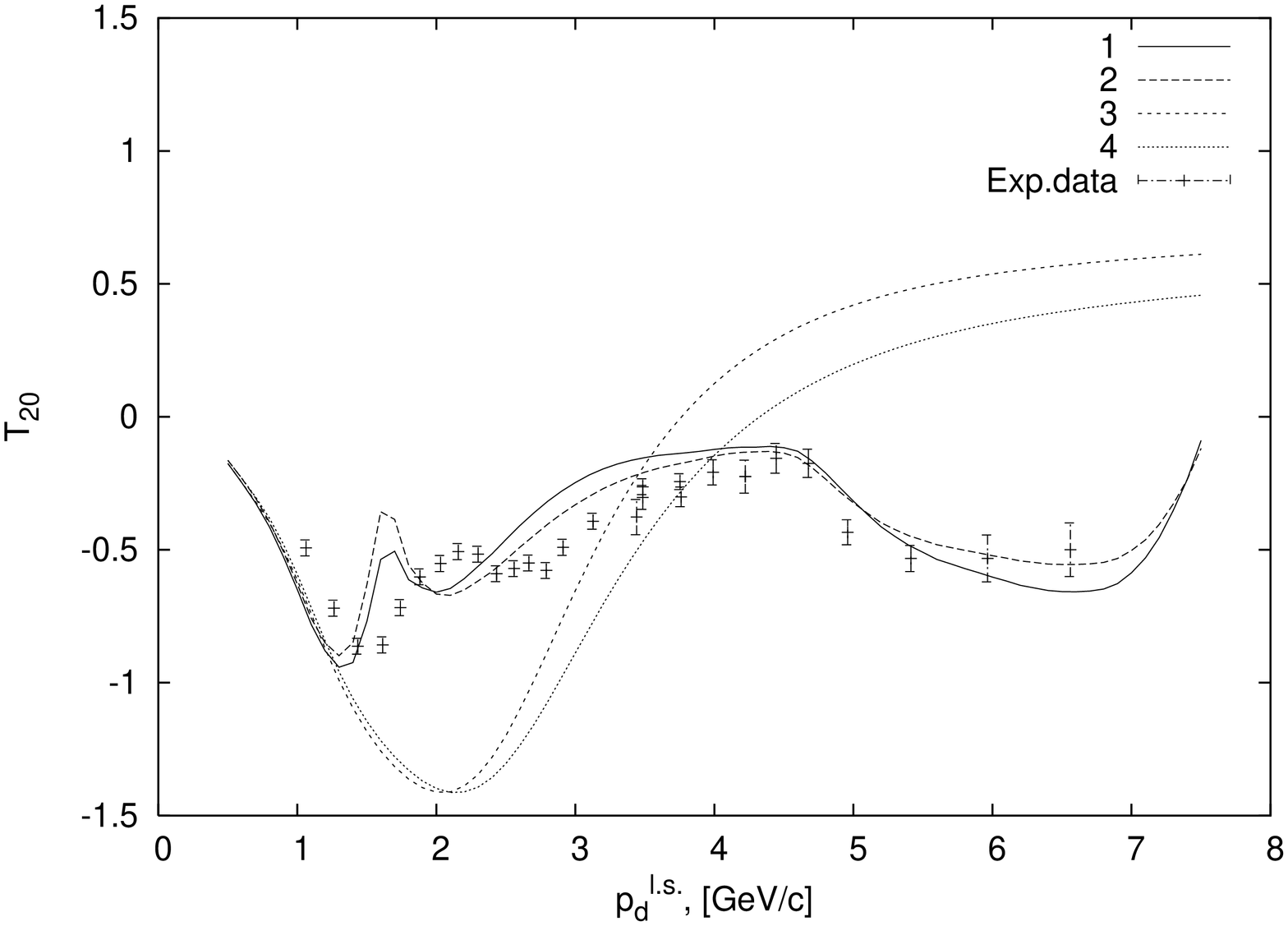, width=65mm,height=65mm,angle=0}}\quad\quad\quad
\rotatebox{270}%
{\psfig{file=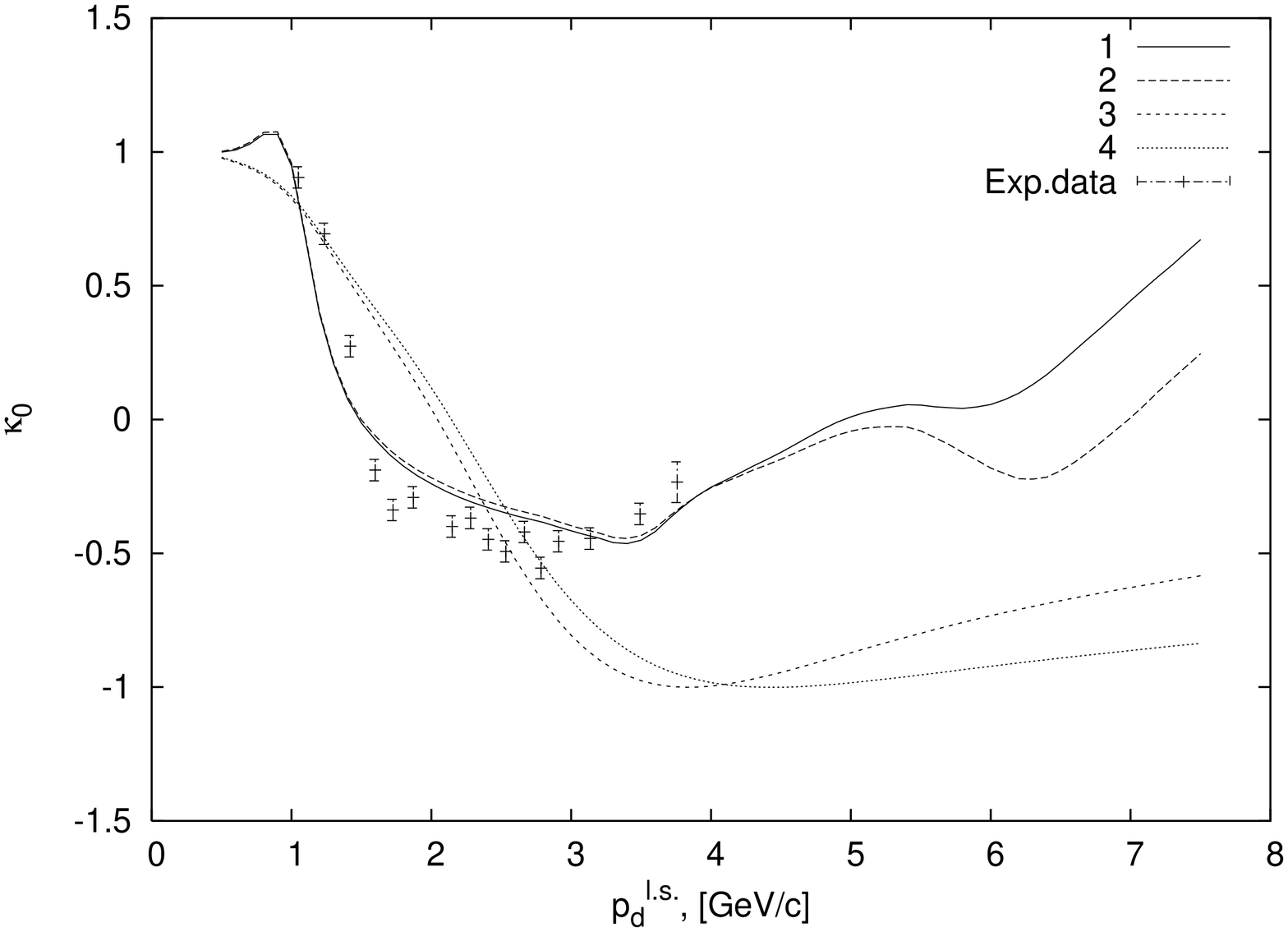,width=65mm,height=65mm,angle=0}}
\caption[Fig.3]{The tensor analyzing power of the deuteron $T_{20}$ as a function of $p_d^{l.s}$ (lhs)
and the transfer polarization $\kappa_0$ as a function of $p_d^{l.s}$ (rhs).}
\end{figure}

{\bf  Acknowledgement.}
We are grateful to F.Gross, E.A.Strokovsky, V.A.Karmanov and Yu.N.Uzikov
for very useful discussions.

\begin{center}
{\bf Discussion}
\end{center}
{\it Question.} {\bf L.N.Strunov.}\\
What can you say about a contribution of one-pion exchange graphs to the deuteron breakup reaction ?\\
{\it Reply.} {\bf G.I.Lykasov.}\\
The contribution of the discussed triangle graphs to the all observables in the deuteron breakup reaction
is sizable at initial energies corresponding to a possible creation of the  $\Delta$-isobar in the
intermediate state, e.g., at the initial kinetic energy about $1~GeV$.\\
{\it Question.} {\bf I.M.Sitnik.}\\
What is a role of the discussed effects in the elastic backward proton-deuteron scattering and the
deuteron stripping reactions on nuclei at high initial energies ? \\
{\it Reply.} {\bf G.I.Lykasov.}\\
`At least, the discussed effects in the elastic backward proton-deuteron scattering decrease 
at initial deuteron momenta above $7~ GeV/c$ and they can be neglected, as is evident from Fig.2.
As for the deuteron stripping reactions on nuclei, probably the contribution of discussed triangle graphs
can be also neglected at high initial energies.\\ 
{\it Question.} {\bf S.L.Belostozky}\\
As I understood, the pion entering into one-pion exchange graphs is virtual. What is the sensitivity of your
results to the pion form factor used in your calculations ?\\
{\it Reply.} {\bf G.I.Lykasov.}\\
We used the monopole form factor for the virtual pion. The sensitivity of all the results to the value
of the cut-off parameter entering into the form factor is about 10-20 percent. The results presented
in the slides correspond to the cut-off parameter about $1~GeV/c$.
{\it Question.} {\bf S.S.Shimansky}\\
Why your old results on $T_{20}$ in the deuteron stripping reaction on a proton including similar 
one-pion exchange graphs did not describe the experimental data at large internal deuteron momenta ?
On the other hand, your new calculations of $T_{20}$ and $\kappa_0$ in the elastic backward proton-deuteron 
scattering allow a rather satisfactory description of the experimental data in the whole kinematic 
region.\\
{\it Reply.} {\bf G.I.Lykasov.}\\
It is due to the following. In our old calculations we did not include the interference between different
graphs, we summed the squares of separate graphs. However, the inclusion of 
the interference terms is very important. Now we include the interference terms because we take the 
pion-nucleon scattering amplitude entering into the $\pi-N$ vertex of the graph in Fig.1c from the  
$\pi-N$ phase shift analysis and can calculate both the real part and the imaginary part of the matrix 
element corresponding to any graph of Fig.1c.  

\end{document}